\documentclass[prd,aps,nofootinbib,singlecolumn]{revtex4-2}
\usepackage{amsmath,amssymb,amsfonts,color,graphicx,graphics,latexsym,placeins,epsfig,multirow}
\usepackage[toc]{appendix}
\usepackage{array}
\newcolumntype{P}[1]{>{\centering\arraybackslash}p{#1}}
\usepackage{footmisc}
\usepackage{mathrsfs}
\usepackage[compatibility=false]{caption}
\usepackage{subcaption}
\usepackage{comment}
\usepackage{bbold}
\usepackage{enumitem}
\setdescription{leftmargin=12.5pt}
\usepackage{bbold}
\usepackage{hyperref}
\usepackage{varioref}
\usepackage{color}

\begin{document}

\title{\Large \bf New formulation of Galilean relativistic Maxwell theory }
\author{Rabin Banerjee \footnote{DAE Raja Ramanna fellow}}
\author{Soumya Bhattacharya}
\affiliation{Department of Astrophysics and High Energy Physics, S.N. Bose National Center for Basic Sciences, Kolkata 700106, India}
\email{rabin@bose.res.in}
\email{soumya557@bose.res.in}


\pdfoutput=1

\newcommand{\sn}{{\rm sn}}
\newcommand{\cn}{{\rm cn}}
\newcommand{\dn}{{\rm dn}}
\newtheorem{theorem}{Theorem}[section]
\newtheorem{lemma}[theorem]{Lemma}

\newtheorem{definition}[theorem]{Definition}
\newtheorem{example}[theorem]{Example}
\newtheorem{xca}[theorem]{Exercise}
\newcommand{\lso}[2]{#1\left(\textcolor{#2}{^{\line(1,0){20}}}\right)}
\newcommand{\lda}[2]{#1\left(\textcolor{#2}{^{_{\bf{------}}}}\right)}
\newcommand{\loo}[2]{#1\left(\textcolor{#2}{^{_{\bf{ooo}}}}\right)}
\newtheorem{remark}[theorem]{Remark}

\def\a{\alpha}
\def\b{\beta}
\def\c{\gamma} 
\def\d{\delta}
\def\e{\epsilon}           
\def\f{\phi}               
\def\vf{\varphi}  \def\tvf{\tilde{\varphi}} 
\def\g{\gamma}
\def\h{\eta}   
\def\i{\iota}
\def\j{\psi}
\def\k{\kappa}                    
\def\l{\lambda}
\def\m{\mu}
\def\n{\nu}
\def\o{\omega}  \def\w{\omega}
\def\p{\pi}                
\def\q{\theta}  \def\th{\theta}                  
\def\r{\rho}                                     
\def\s{\sigma}                                   
\def\t{\tau}
\def\u{\upsilon}
\def\x{\xi}
\def\z{\zeta}
\def\D{\Delta}
\def\F{\Phi}
\def\G{\Gamma}
\def\J{\Psi}
\def\L{\Lambda}
\def\O{\Omega}  \def\W{\Omega}
\def\P{\Pi}
\def\Q{\Theta}
\def\S{\Sigma}
\def\U{\Upsilon}
\def\X{\Xi}
\def\del{\partial}              


\def\ca{{\cal A}}
\def\cb{{\cal B}}
\def\cc{{\cal C}}
\def\cd{{\cal D}}
\def\ce{{\cal E}}
\def\cf{{\cal F}}
\def\cg{{\cal G}}
\def\ch{{\cal H}}
\def\ci{{\cal I}}
\def\cj{{\cal J}}
\def\ck{{\cal K}}
\def\cl{{\cal L}}
\def\cm{{\cal M}}
\def\cn{{\cal N}}
\def\co{{\cal O}}
\def\cp{{\cal P}}
\def\cq{{\cal Q}}
\def\car{{\cal R}}
\def\cs{{\cal S}}
\def\ct{{\cal T}}
\def\cu{{\cal U}}
\def\cv{{\cal V}}
\def\cw{{\cal W}}
\def\cx{{\cal X}}
\def\cy{{\cal Y}}
\def\cz{{\cal Z}}



\def\6{\partial}
\vskip .2in
\begin{abstract}

In this paper, we discuss Galilean relativistic Maxwell theory in detail. We first provide a set of mapping relations, derived systematically, that connect the covariant and contravariant vectors in the Lorentz relativistic and Galilean relativistic formulations. Exploiting this map, we construct the two limits of Galilean relativistic Maxwell theory from usual Maxwell's theory in the potential formalism for both contravariant and covariant vectors which are now distinct entities. Field equations are derived and their internal consistency is shown. The entire analysis is then performed in terms of electric and magnetic fields for both covariant and contravariant components. Duality transformations and their connection with boost symmetry are discussed which reveal a rich structure. The notion of twisted duality is introduced. Next we consider  gauge symmetry, construct Noether currents and show their on-shell conservation. We also discuss shift symmetry under which the Lagrangian is invariant, where the corresponding currents are now on-shell conserved. At the end we analyse the theory by including sources for both contravariant and covariant sectors. We show that sources are now  off-shell conserved
\end{abstract}

\maketitle
\section{Introduction}
\noindent The formulation of non-relativistic limit of classical field theories received considerable attention recently. It has found applications in  holography \cite{Taylor}, in studying non-relativistic diffeomorphisms (NRDI) \cite{andreev1, andreev2, jensen, rb1, rb2}, 
condensed matter systems \cite{Son, Pal, Geracie}, fluid dynamics \cite{Jain, rb3}, gravitation \cite{Morand, Read}. This formulation is tricky and markedly different from the relativistic
case. Covariance in non-relativistic physics is subtle due to the absolute nature of time. The lack of a single non-degenerate metric in the non-relativistic limit poses some additional difficulties. Here we are interested in the non-relativistic limit of  Maxwellean electrodynamics which is invariant under galilean transformations. The basic construction of Galilean electrodynamics was first given by Le Bellac and Levy-Leblond \cite{Leblond} back in 1970's. This was done in the field formulation. A similar field based analysis was done in \cite{Khanna} using embedding techniques. Further directions in this type of analysis were provided in \cite{Duval}. Other references on different aspects of Galilean electrodynamics and gauge theories are \cite{Mehra1,  Bleeken, Bergshoeff, Festuccia, Mehra3, Chapman, Sharma}. \\
\indent In this paper we provide a detailed analysis of galilean relativistic Maxwell theory with and without sources. While earlier findings are reproduced we also find several new results with new interpretations. We know there are two distinct non-relativistic limits possible for electrodynamics known as electric and magnetic limit \cite{Leblond, Duval}. We derive these two limits from the Lorentz transformation of an arbitrary four vector. Our derivation of the non-relativistic scaling relations  are consistent with \cite{Leblond, Duval}. All previous works so far treated only the contravariant components of any vector quantity but the novelty of our treatment is that we have considered both contravariant and covariant components separately (which are distinct quantities in the non-relativistic case) and hence help us to explore the rich symmetries involved in the theory. We then derive the Lagrangians for both electric and magnetic limits from which the equations of motion are obtained. We also show that Maxwell's equations under non-relativistic limit (electric and magnetic) yield same equations as those we get from the non-relativistic Lagrangians. This implies the internal consistency of the limiting process. We observe that at the equations of motion level if we replace the covariant vector components by the corresponding contravariant ones then the electric limit and magnetic limits get interchanged. We then define the galilean electric and magnetic fields for contravariant and covariant cases.  Along the way, we discuss the dualities  and point out some of the subtleties involved in the process. In this regard we observe one unique aspect of the duality symmetries called {\it twisted duality} which is valid only in the galilean limit. Especially we show that the transformations of the electric and magnetic fields under galilean boosts is connected with the familiar duality transformations. Next we move to discuss gauge symmetry. We have shown that we can choose different gauge parameters for contravariant and covariant four potentials as they represent different entities in the galilean limits. We then compute the galilean version of the Noether currents and explicitly show their on-shell conservation. We discuss shift symmetries which play an important role in the study of low-energy effective lagrangians in the context of Goldstone's theorem. Recently people have explored shift symmetries from different aspects \cite{rb4, rb5}. We compute corresponding currents and their conservations in this limit. In the end we introduce sources and write down the Lagrangians for appropriate galilean limits (electric and magnetic) and the equations of motion just like the sourceless case. \\
\noindent The paper is organised as follows: in section \ref{sec2} we derive mapping relations between relativistic and non-relativistic vectors for electric and magnetic limit for both contravariant and covariant vectors. In section \ref{sec3} we derive the non-relativistic lagrangian for both limits and write down the equations of motion. We discuss Maxwell's equations in terms of fields and explore the duality relations in \ref{sec4}. Gauge symmetry, Noether currents and their conservations are discussed in \ref{sec5}. In section \ref{sec6} we discuss  shift symmetry and its galilean counterpart, corresponding currents and their conservations. In section \ref{sec7}, discussion has been done by including sources for both contravariant and covariant sectors.  Finally, conclusions have been given in \ref{sec8}. 
\section{ Mapping relations} \label{sec2}
\noindent Here we derive a certain scaling between special relativistic and Galilean relativistic quantities. As we know there exists two types of such limits for the vector quantities namely electric and magnetic limits. So first let us consider the contravariant vectors. Let us consider a generic Lorentz transformation with the boost velocity as $u^i$:
\begin{equation}
    x'^0 = \gamma x^0 - \frac{\gamma u_i}{c} x^i
    \label{t1}
\end{equation}
\begin{equation}
    x'^i = x^i - \frac{\gamma u^i}{c}x^0 + (\gamma -1 )\frac{u^i u_j}{u^2}x^j
    \label{t2}
\end{equation}
where $\gamma = \frac{1}{\sqrt{1-\frac{u^2}{c^2}}}$. Under such Lorentz transformations a contravariant vector changes as
\begin{equation*}
    V'^{\mu} = \frac{\6 x'^{\mu}}{\6 x^{\nu}} V^{\nu}
\end{equation*}
We can write them component-wise as (also considering $u<<c$, so $\gamma \to 1$)
\begin{equation}
    V'^0 = V^0 - \frac{u_j}{c}V^j
    \label{contra1}
\end{equation}
\begin{equation}
    V'^i = V^i - \frac{u^i}{c} V^0
    \label{contra2}
\end{equation}
We next provide a map that relates the Lorentz vectors with their Galilean counterparts. 
\footnote{Notation: Here relativistic vectors are denoted by capital letters ($V^0, V^i$ etc) and Galilean vectors are denoted by lowercase letters ($v^0, v^i$ etc).} 
\begin{equation}
    V^0 = c v^0, \,\,\,\, V^i = v^i
    \label{contrael}
\end{equation}
This particular map corresponds to the case $\frac{V^0}{V^i} = c ~\frac{v^0}{v^i}$ in the $c \to \infty$ limit. This yields largely timelike vectors and is called 'electric limit'.
Now using eqn \ref{contrael} in eqns \ref{contra1} and \ref{contra2} we get
\begin{equation}
v'^0 = v^0
\label{v1}
\end{equation}
\begin{equation}
    v'^i = v^i - u^i v^0
    \label{v2}
\end{equation}
The above two equations define the Galilean transformations. We can write them in a single matrix equation as
\begin{equation}
\begin{pmatrix}
v'^0 \\ v'^i
\end{pmatrix} 
= \begin{pmatrix}
1 & 0\\ -u^i & 1
\end{pmatrix}
\begin{pmatrix}
v^0 \\ v^i
\end{pmatrix}
\label{V1}
\end{equation}
We now consider the magnetic limit which corresponds to largely spacelike vectors
\begin{equation}
    V^0 = -\frac{v^0}{c}, \,\,\,\, V^i = v^i
    \label{contramag}
\end{equation}
Now using \ref{contramag} in \ref{contra1} and \ref{contra2} we get

\begin{equation}
    v'^0 = v^0 + u_j v^j
    \label{v3}
\end{equation}
\begin{equation}
    v'^i = v^i
    \label{v4}
\end{equation}
which is again a galilean transformation. We can write eqn \ref{v3} and \ref{v4} as a matrix equation 
\begin{equation}
  \begin{pmatrix}
  v'^0 \\ v'^i 
  \end{pmatrix} 
  = \begin{pmatrix}
  1 & u_j \\ 0 & 1
  \end{pmatrix}
  \begin{pmatrix}
  v^0 \\ v^j
  \end{pmatrix}
  \label{V2}
\end{equation}
We will now consider the covariant vectors. We will write first the reverse transformations of eqn \ref{t1} and \ref{t2} which is 
\begin{equation}
    x^0 = \gamma x'^0 + \frac{\gamma u_i}{c} x'^i
    \label{t3}
\end{equation}
\begin{equation}
    x^i = x'^i + \frac{\gamma u^i}{c}x'^0 + (\gamma -1 )\frac{u^i u_j}{u^2}x'^j
    \label{t4}
\end{equation}
And we know covariant vectors transform as
\begin{equation*}
    V'_{\mu} = \frac{\6 x^{\nu}}{\6 x'^{\mu}} V_{\nu}
\end{equation*}
Componentwise we can again write them as
\begin{equation}
    V'_0 = V_0 + \frac{u^i}{c} V_i
    \label{cov1}
\end{equation}
\begin{equation}
    V'_i = V_i + \frac{u_i}{c} V_0
    \label{cov2}
\end{equation}
Now here we take the  electric limit in the following way, which will soon become clear
\begin{equation}
    V_0 = \frac{v_0}{c}, \,\,\,\, V_i = v_i
    \label{covel}
\end{equation}
Using \ref{covel} in \ref{cov1} and \ref{cov2} we get
\begin{eqnarray}
  v'_0 = v_0 + u^i v_i
  \label{v5}
\end{eqnarray}
\begin{eqnarray}
  v'_i = v_i
  \label{v6}
\end{eqnarray}
which are again Galilean transformations. We can write \ref{v5} and \ref{v6} as a matrix equation as
\begin{equation}
 \begin{pmatrix}
 v'_0 \\ v'_i 
 \end{pmatrix} 
 = \begin{pmatrix}
 1 & u_i \\ 0 & 1
 \end{pmatrix}
 \begin{pmatrix}
 v_0 \\ v_i
 \end{pmatrix}
 \label{V3}
\end{equation}
We will now consider the magnetic limit as 
\begin{equation}
    V_0 = -c v_0, \,\,\,\, V_i = v_i
    \label{covmag}
\end{equation}
Using \ref{covmag} in \ref{cov1} and \ref{cov2} we get 
\begin{equation}
    v_0' = v_0
    \label{v7}
\end{equation}
\begin{equation}
    v'_i = v_i - u_i v_0 
    \label{v8}
\end{equation}
We can write \ref{v7} and \ref{v8} as 
\begin{equation}
    \begin{pmatrix}
    v'_0 \\ v'_i 
    \end{pmatrix}
    = \begin{pmatrix}
    1 & 0 \\ -u_i & 1 
    \end{pmatrix}
    \begin{pmatrix}
    v_0 \\ v_i
    \end{pmatrix}
    \label{V4}
\end{equation}
We can show that the transformation matrix in \ref{V1} and the transpose of the matrix \ref{V3} satisfies
\begin{equation}
 \begin{pmatrix}
1 & 0\\ -v^i & 1
\end{pmatrix}
 \begin{pmatrix}
1 & 0\\ v^i & 1
\end{pmatrix} = \begin{pmatrix}
1 & 0 \\ 0 & 1
\end{pmatrix}
\end{equation}
Similarly the transformation matrix in equation \ref{V2} and the transpose of the transformation matrix in \ref{V4} satisfies
\begin{equation}
 \begin{pmatrix}
  1 & v_j \\ 0 & 1
  \end{pmatrix} 
  \begin{pmatrix}
   1 & -v_j \\ 0 & 1
  \end{pmatrix} = \begin{pmatrix}
   1 & 0 \\ 0 & 1
  \end{pmatrix}
\end{equation}

\noindent To justify the limiting prescriptions even further, we consider the norm preservation for both electric and magnetic limits. Let us first consider the norm in the electric limit
\begin{equation}
    V^0 V_0 + V^i V_i \xrightarrow[\text{limit}]{\text{electric}} \Big(c v^0 \Big)\Big(\frac{v_0}{c} \Big) + \Big(v^i\Big) \Big( v_i\Big) = v^0 v_0 + v^i v_i
\end{equation}
which clearly indicates that under the scaling, the norm is preserved. Likewise, the norm in the magnetic limit is also conserved. \\
\begin{table}
\caption{Mapping relations}\label{T1}
\begin{center}
\begin{tabular}{|c|c|c|} \hline 
${\rm Limit}$  & $ {\rm Contravariant~~mapping}$ & $ {\rm Covariant~~mapping}$  \\ \hline
${\rm Electric ~~limit}$ & $V^0 \to c~v^0, \,\,  V^i \to v^i$ & $V_0 \to \frac{v_0}{c}, \,\, V_i \to v_i$ \\ \hline
${\rm Magnetic ~~limit}$ & $V^0 \to -\frac{v^0}{c},\,\,V^i \to v^i $ & $V_0 \to -c~v_0. \,\, V_i \to v_i$ \\
\hline
\end{tabular}
\label{T1}
\end{center}
\end{table}
\noindent The mapping relations, systematically derived here for both covariant and contravariant components, are essential to the subsequent analysis. Any four vector in relativistic theory will be replaced by the corresponding structure for the Galilean theory by adopting this map. These relations are summarised in table \ref{T1}.
\section{Lagrangian and field equations} \label{sec3}
\noindent Now let us start from the relativistic Maxwell theory described by the Lagrangian
\begin{equation}
    \mathcal{L} = -\frac{1}{4}~F_{\mu \nu}~F^{\mu \nu} = -\frac{1}{4} \eta_{\mu \alpha} \eta_{\nu \beta} F^{\alpha \beta} F^{\mu \nu} \label{maxwll}
\end{equation}
where $F_{\mu \nu} = \partial_{\mu} A_{\nu} - \partial_{\nu} A_{\mu}$ and $\eta_{\mu \nu}$ is the flat space metric with signature $\Big(-,+,+,+\Big)$. 
\subsection{Electric limit}
Now using the relations given in table \ref{T1} we can write the two terms in \ref{maxwll} as, 
\begin{eqnarray}
2 F_{0i}F^{0i} 
\xrightarrow[\text{$c \to \infty$}]{\text{Electric limit}}  -2 \6^i a^0 \Big( \partial_t a_i - \partial_i a_0  \big) \hspace{1in}
\label{part1}
\end{eqnarray}
 
\begin{eqnarray}
F_{ij}F^{ij}  \xrightarrow[\text{$c \to \infty$}]{\text{Electric limit}} \Big(\6_i a_j - \6_j a_i \Big) \Big( \6^i a^j - \6^j a^i \Big) \equiv f_{ij} f^{ij}
\end{eqnarray}

\noindent Here $A^{\mu}$ is the relativistic four potential while $a^0$ and $a^i$ are it's galilean counterpart. So in the electric limit the full lagrangian takes the following form
\begin{equation}
    \mathcal{L}_e = \frac{1}{2} \6^i a^0 \Big( \partial_t a_i - \partial_i a_0  \big)  - \frac{1}{4} f_{ij} f^{ij}
    \label{el1}
\end{equation}



\noindent Now we derive the equations of motion. Varying the Lagrangian \ref{el1} with respect to $a_0,~a_j,~a^0,~a^j$ we get the corresponding equations of motion
\begin{eqnarray}
\6_i \6^i a^0 = 0 
\label{ee1}
\end{eqnarray}
\begin{eqnarray}
\6_t \6^j a^0 + \6_i \6^j a^i - \6_i \6^i a^j = 0
\label{ee2}
\end{eqnarray}
\begin{eqnarray}
\6^i \6_t a_i - \6^i \6_i a_0 = 0
\label{ee3}
\end{eqnarray}
\begin{eqnarray}
\6^i \6_i a_j - \6^i \6_j a_i = 0 
\label{ee4}
\end{eqnarray}
We now derive these equations directly from the equations of motion. The relativistic equations, in component form, are given by,
\begin{eqnarray}
\6_i F^{i0} =0,\,\,\,\,\,\,~~~~~~\6_0 F^{0j} + \6_i F^{ij} = 0 \label{e2}
\end{eqnarray}
The Galilean version of these equations is found by using table \ref{T1}, followed by taking $c \to \infty$. It reproduces \ref{ee1} and \ref{ee2} respectively. 
To get the remaining pair of equations we have to interpret the relativistic  eqns given in \ref{e2} as
\begin{eqnarray}
\6^i F_{i0} =0, \,\,\,\,\,~~~~~~~~~\6^0 F_{0j} + \6^i F_{ij} = 0 \label{e3}
\end{eqnarray}
Once again the Galilean version is obtained from table \ref{T1}, followed by taking $c \to \infty$. Equations \ref{ee3}, \ref{ee4} are reproduced. 
This shows the consistency of the eqn of motion in Galilean electrodynamics.
\subsection{Magnetic limit}
Here again using the relations given in table \ref{T1} we can write the two terms in \ref{maxwll} as,
\begin{eqnarray}
2 F_{0i}F^{0i} \xrightarrow[\text{$c \to \infty$}]{\text{Magnetic limit}} -2 \6_i a_0 \Big(\6_t a^i - \6^i a^0 \Big) \hspace{1in}
\end{eqnarray}
\begin{eqnarray}
F_{ij}F^{ij}  \xrightarrow[\text{$c \to \infty$}]{\text{Magnetic limit}} \Big(\partial_i a_j - \partial_j a_i \Big) \Big(\partial^i a^j - \partial^j a^i \Big)
\equiv f_{ij}f^{ij} 
\end{eqnarray}
So the Lagrangian will take the following form

\begin{equation}
\mathcal{L}_m = \frac{1}{2}  \6_i a_0 \Big(\6_t a^i - \6^i a^0 \Big) - \frac{1}{4} f_{ij}f^{ij} \label{l2}
\end{equation}
Varying \ref{l2} wrt $a_0,~a_j,~a^0,~a^j$ we get,
\begin{eqnarray}
   \partial_i ~\partial_t~ a^i - \partial^i ~\partial_i ~a^0 = 0 
  \label{em1}
\end{eqnarray}
\begin{eqnarray}
   \6_j\6^i a^j - \6_j \6^j a^i = 0 
  \label{em2}
  \end{eqnarray}
\begin{eqnarray}
  \6^i \6_i a_0 = 0
  \label{em3}
\end{eqnarray}
\begin{eqnarray}
 \6_t \6_j a_0 + \6^i \6_j a_i - \6^i \6_i a_j = 0 \label{em4}
\end{eqnarray}
Here also we can show that the above equations agree with those derived directly from relativistic Maxwell equations \ref{e2} and \ref{e3} corresponding to contravariant and covariant sectors respectively, by taking the magnetic limit.

\begin{table}
\caption{Field equations}\label{T2}
\begin{center}
\begin{tabular}{|c|c|c|} \hline 
${\rm Variables}$ & ${\rm Electric ~~limit}$ & ${\rm Magnetic ~~limit}$ \\ \hline
$a^0$  & $\6^i \6_t a_i - \6^i \6_i a_0 = 0$ & $\6^i \6_i a_0 = 0$ \\ \hline
$a^i$ & $\6^j \6_i a_j - \6^j \6_j a_i = 0 $ & $\6_t \6_i a_0 + \6^j \6_i a_j - \6^j \6_j a_i = 0$ \\ \hline
$a_0$ & $ \partial^i \partial_i a^0 = 0 $ & $\partial_i ~\partial_t~ a^i - \partial^i ~\partial_i ~a^0 = 0 $ \\ \hline
$a_i$ & $\partial_t \partial^i a^0 + \partial_j \partial ^i a ^j - \partial_j \partial^j a^i = 0$ & 
$\6_j\6^i a^j - \6_j \6^j a^i = 0 $ \\ \hline
\end{tabular}
\label{T2}
\end{center}
\end{table}
\noindent The field equations for both the limits of Galilean electrodynamics are shown in table \ref{T2}.
\section{Galilean electric and magnetic fields and dual transformations} \label{sec4}
\noindent Here we introduce the galilean limit of electric and magnetic fields and write down the Maxwell equations. For this purpose we will discuss Contravariant and Covariant sector separately.
\subsection{Contravariant sector}
Relativistic electric and magnetic fields are defined as 
\begin{eqnarray}
E^i = \6^0 A^i - \6^i A^0 \\
B^i = \epsilon^{ij}_{~k} \6_j A^k
\end{eqnarray}
First, we consider the electric limit. \\
\noindent \underline {\bf Electric limit:}\\

Using the mapping relations given in table \ref{T1} we can write the electric field as 
\begin{eqnarray}
E^i = -\frac{1}{c} \6_t a^i - c \6^i a^0 
\end{eqnarray}
And we can define the Galilean electric and magnetic fields as 
\begin{equation}
    e^i = \lim_{c \to \infty} \frac{E^i}{c} = -\6^i a^0. \,\,\,~~~~~ b^i =  \lim_{c \to \infty} B^i = \epsilon^{ij}_{~k} \6_j a^k \label{emapping1}
\end{equation}

\noindent    Now we write the field equations that we derived in the previous section  in terms of the Galilean electric and magnetic fields. From \ref{ee1} we get 
    \begin{eqnarray}
    \6_i \6^i a^0 = 0 \implies \6_i (-e^i) = 0 \implies \vec \nabla . \vec e = 0 
    \end{eqnarray}
\noindent    Similarly eqn \ref{ee2} implies
    \begin{eqnarray}
   \6_t \6^j a^0 + \6_i \6^j a^i - \6_i \6^i a^j = 0 
     \implies  (\vec \nabla \times \vec b)^j = \6_t e^j
    \end{eqnarray}
    
\noindent We can see clearly that
\begin{equation}
 \vec \nabla .\vec b =  \6_i b^i = \6_i \epsilon^{ij}_{~k} \6_j a^k =\epsilon^{ij}_{~k} \6_i\6_j a^k = 0  
\end{equation}

\noindent We will now compute $\vec \nabla \times \vec e$,
\begin{eqnarray}
(\nabla \times e)^i = \epsilon^{ij}_{~k} \6_j e^k 
= \epsilon^{ij}_{~k} \6_j (-\6^k a^0) = 0 
\end{eqnarray}
So in electric limit we get the following set of equations
\begin{eqnarray}
\vec \nabla . \vec e = \6_i e^i = 0, \label{max01}\\
\vec \nabla . \vec b =\6_i b^i = 0, \label{max02}\\ 
(\vec \nabla \times \vec e)^i = \epsilon^{ij}_{~k} \6_j e^k = 0, \label{max1} \\
(\vec \nabla \times \vec b)^i = \epsilon^{ij}_{~k} \6_j b^k = \6_t (\vec e)^i \label{max2}
\end{eqnarray}
\noindent It is now possible to obtain the equations (\ref{max01} - \ref{max2}) directly from Maxwell's equations,
\begin{equation}
    \vec \nabla.\vec E = 0, ~~~~\vec \nabla.\vec B = 0,~~~~\vec \nabla \times \vec E = -\frac{1}{c} \frac{\6 \vec B}{\6 t}, ~~~~\vec \nabla \times \vec B = \frac{1}{c} \frac{\6 \vec E}{\6 t} \label{MAX}
\end{equation}
by using the identification in eqn \ref{emapping1} and taking $c \to \infty$.\\
\indent For the electric limit, it is seen from \ref{max2}, a change in the electric field influences the magnetic field. But a change in the magnetic field does not influence the electric field since the R.H.S of \ref{max1} vanishes. This implies that the electric field is considerably greater and dominates over the magnetic field, justifying the nomenclature {\it electric limit}.  This is different from the relativistic case where electric and magnetic fields are treated symmetrically. This asymmetry in Galilean electromagnetism leads to physical effects, some of which have been discussed in \cite{Germain}.\\
\indent We will now consider the magnetic limit.\\
\noindent \underline {\bf Magnetic limit:}\\

Electric field can be written in this limit from the mapping relations in \ref{T1} as 
\begin{equation}
    E^i = -\frac{1}{c} \6_t a^i + \frac{1}{c} \6_i a^0
\end{equation}
And we can define the galilean electric and magnetic fields as 
\begin{equation}
    e^i = \lim_{c \to \infty} c E^i = -(\6_t a^i - \6^i a^0), \,\,\, ~~~~~ b^i = \lim_{c \to \infty} B^i = \epsilon^{ij}_{~k} \6_j a^k \label{bmapping1}
\end{equation}
Using these relations, eqns \ref{em1}, \ref{em2} and two Bianchi identities are expressed in terms of electric/magnetic fields as
\begin{eqnarray}
\vec \nabla . \vec e = \6_i e^i = 0, \label{max03}\\
\vec \nabla . \vec b =\6_i b^i = 0, \label{max04}\\ 
(\vec \nabla \times \vec e)^i = \epsilon^{ij}_{~k} \6_j e^k = -\6_t (\vec b)^i, \label{max3} \\
(\vec \nabla \times \vec b)^i = \epsilon^{ij}_{~k} \6_j b^k  = 0 \label{max4}
\end{eqnarray}
The above equations (\ref{max03} - \ref{max4}) also follow from Maxwell's equations \ref{MAX}, using the map \ref{bmapping1} and taking the limit $c \to \infty$. In contrast to the electric limit, here a change in the magnetic field influences the electric field but the converse does not hold. In this case the magnetic field dominates over the electric field.\\
\indent We can clearly see that equations in electric limit are mapped to those of magnetic limit and vice  versa under the following duality transformations,
\begin{equation}
    e^i \to b^i, \, \, ~~~~ b^i \to -e^i 
\end{equation}
\begin{equation}
    e^i \to -b^i, \, \, ~~~~ b^i \to e^i 
\end{equation}
\noindent This is the analogue of the electromagnetic duality in usual Maxwell's source free theory. 
\begin{table}
\caption{Fields in galilean limit}\label{T05}
\begin{center}
\begin{tabular}{|c|c|c|} \hline 
${\rm Limits}$  & $ {\rm Electric ~field}$ & $ {\rm Magnetic ~field}$  \\ \hline
${\rm Electric ~limit}$ & $E^i \to c e^i, ~E_i \to \frac{e_i}{c}$ & $B^i \to b^i, ~B_i \to b_i$ \\ \hline
${\rm Magnetic ~limit }$ &  $ E^i \to \frac{e^i}{c}, ~E_i \to c e_i$  & $B^i \to b^i, ~B_i \to b_i $ \\
\hline
\end{tabular}
\label{T05}
\end{center}
\end{table}
\subsection{Covariant sector}
Relativistic electric and the magnetic fields are defined as 
\begin{eqnarray}
E_i = - \Big( \6_0 A_i - \6_i A_0\Big) \\
B_i = \epsilon_i^{~jk} \6_j a_k
\end{eqnarray}
Now we consider the electric limit. \\
\noindent \underline {\bf Electric limit:}\\
In this limit the electric field looks like 
\begin{equation}
E_i = -\Big(\frac{1}{c} \6_t a_i + \frac{1}{c} \6_i   a_0\Big)
\end{equation}
And we can define Galilean electric and magnetic fields as 
\begin{equation}
    e_i = \lim_{c \to \infty} c E_i = - (\6_t a_i - \6_i a_0), \,\,\,~~~~~~ b_i = \lim_{c \to \infty} B_i = \epsilon_{i}^{~jk} \6_j a_k   \label{emapping2}
\end{equation}

From eqn \ref{ee3} we get
\begin{eqnarray}
 \6^i \Big( \6_t a_i - \6_i a_0 \Big)
= \6^i (-e_i) = \6^i e_i = \vec \nabla . \vec e =0 
\end{eqnarray}
Eqn \ref{ee4} yields
\begin{eqnarray}
\6^i f_{ij}
= -\epsilon_{ji}^{~~k}\6^i b_k =0 \implies \Big(\vec \nabla \times \vec b \Big)_j = 0
\end{eqnarray}
And finally calculation of $\vec \nabla \times \vec e$ yields,
\begin{eqnarray}
\Big(\vec \nabla \times \vec e \Big)_i = \epsilon_{i}^{~jk} \6_j e_k  = -\epsilon_{i}^{jk}\6_j \Big(\6_t a_k - \6_k a_0  \Big)
= -\6_t \epsilon_{i}^{~jk} \6_j a_k  = - \6_t b_i 
\end{eqnarray}
So the Maxwell equations in the electric limit are 
\begin{eqnarray}
\vec \nabla . \vec e = \6^i e_i = 0, \label{max05}\\
\vec \nabla . \vec b = \6^i b_i = 0, \label{max06}\\ 
\Big(\vec \nabla \times \vec e\Big)_i = \epsilon_{i}^{~jk}\6_j e_k = - \6_t (\vec b)_i, \label{max5} \\
\Big(\vec \nabla \times \vec b \Big)_i= \epsilon_{ij}^{~~k} \6^j b_k = 0  \label{max6}
\end{eqnarray}
The above equations (\ref{max05} - \ref{max6}) also follow from Maxwell's equations \ref{MAX}, using the map \ref{emapping2} and taking the limit $c \to \infty$.
\noindent \underline {\bf Magnetic limit:}\\
\begin{equation*}
    A_0 \to -c a_0 \, \, \, \, A_i \to a_i 
\end{equation*}
In this limit electric field is scaled as 
\begin{equation}
    E_i = -\Big(\frac{1}{c} \6_t a_i + c \6_i a_0 \Big)
\end{equation}
And we can define the Galilean electric and magnetic fields as 
\begin{equation}
    e_i = \lim_{c \to \infty} \frac{E_i}{c} = -\6_i a_0, \,\,\, ~~~~~ b_i = \lim_{c \to \infty} B_i = \epsilon_{i}^{~jk} \6_j a_k  \label{bmapping2}
\end{equation}

   The equations \ref{em3}, \ref{em4} and two Bianchi identities are now written as 
\begin{eqnarray}
\vec \nabla . \vec e = \6^i e_i = 0, \label{max07}\\
\vec \nabla . \vec b = \6^i b_i = 0, \label{max08}\\ 
\Big(\vec \nabla \times \vec e \Big) = \epsilon_{i}^{~jk} \6_j e_k = 0, \label{max7} \\
\Big(\vec \nabla \times \vec b \Big)_i = \epsilon_{ij}^{~~k} \6^j b_k = \6_t e_i \label{max8}
\end{eqnarray}
The above equations (\ref{max07} - \ref{max8}) also follow from Maxwell's equations \ref{MAX}, using the map \ref{bmapping2} and taking the limit $c \to \infty$.
Here also we see that electric and magnetic fields satisfy certain duality relations as follows 
\begin{equation}
    e_i \to b_i, \, \, ~~~~ b_i \to -e_i 
\end{equation}
\begin{equation}
    e_i \to -b_i, \, \, ~~~~ b_i \to e_i 
\end{equation}
The galilean limit scalings for the fields (electric/magnetic) are shown in Table \ref{T05}.
\subsection{Effect of the dualities at the level of Lagrangian}
\noindent In the electric limit the Lagrangian is represented by eqn \ref{el1}. 
Now in this limit, the contravariant and covariant electric fields as well as magnetic fields are represented by \ref{emapping1} and \ref{emapping2} respectively.
Using these definitions we can write the electric limit Lagrangian in the following form
\begin{equation}
    \mathcal{L}_e =  \frac{1}{2} \Big(e^i e_i - b_i b^i \Big) \label{lfe}
\end{equation}
\noindent Similarly in the magnetic limit the Lagrangian is represented by eqn \ref{l2}. 
Here electric and magnetic fields for contravariant and covaraint cases are given in equations \ref{bmapping1} and \ref{bmapping2} respectively. 
So now the Lagrangian in this limit takes the following form in terms of the fields
\begin{equation}
    \mathcal{L}_m =  \frac{1}{2} \Big(e^i e_i - b_i b^i \Big) \label{lfm}
\end{equation}

\noindent We observe that both Lagrangians (eqn \ref{lfe} and \ref{lfm}) are identical  
\begin{equation}
    \mathcal{L}_e = \mathcal{L}_m = \mathcal{L}
\end{equation}
In other words, expressed in terms of the gauge invariant fields (electric and magnetic), the lagrangians in the two limits are same. This is to be contrasted with the potential formulation where $\mathcal{L}_e$ and $\mathcal{L}_m$ appear to be different. However if we interchange the covariant and the contravariant indices then $\mathcal{L}_e$ and $\mathcal{L}_m$ get interchanged. Similar things happen when the lagrangians are expressed in terms of the electric and magnetic fields. However, since the expressions are symmetrical with respect to the covariant and contravariant indices, $\mathcal{L}_e$ and $\mathcal{L}_m$ become identical.\\
\noindent We observe there is an overall sign change (i.e $\mathcal{L} \to -\mathcal{L}$) under the duality transformations ($e^i \to b^i, ~ b^i \to -e^i, ~e_i \to b_i, ~ b_i \to -e_i$ or $e^i \to -b^i, ~b^i \to e^i, ~e_i \to -b_i, ~b_i \to e_i$) however the Lagrangians remain invariant (i.e $\mathcal{L} \to \mathcal{L}$) under the twisted duality relations (i.e $e^i \to -b^i, ~ b^i \to e^i, ~e_i \to b_i, ~ b_i \to -e_i$ or $e^i \to b^i, ~ b^i \to -e^i, ~e_i \to -b_i, ~ b_i \to e_i$).  This has been shown clearly in table \ref{T03}. We like to mention that the twisted relations have not been discussed earlier.
\begin{table}
\caption{Effect of duality on the Lagrangian}\label{T03}
\begin{center}
\begin{tabular}{|c|c|} \hline 
${\rm Duality ~relation}$  & $ {\rm Change ~in ~the ~Lagrangian}$  \\ \hline
$e^i \to b^i, ~b^i \to -e^i, ~e_i \to b_i, ~b_i \to -e_i$ & $\mathcal{L} \to -\mathcal{L}$ \\ \hline
$e^i \to -b^i, ~b^i \to e^i, ~e_i \to -b_i, ~b_i \to e_i$ & $\mathcal{L} \to -\mathcal{L} $ \\ \hline
$e^i \to b^i, ~b^i \to -e^i, ~e_i \to -b_i, ~b_i \to e_i$ & $\mathcal{L} \to \mathcal{L} $ \\ \hline
$e^i \to -b^i, ~b^i \to e^i, ~e_i \to b_i, ~b_i \to -e_i$ & $\mathcal{L} \to \mathcal{L} $ \\ 
\hline
\end{tabular}
\label{T03}
\end{center}
\end{table}

\subsection{Dual transformation of electric and magnetic fields under Galilean boost}
\noindent \underline{\bf Contravariant case:} \\
\noindent The Field transforms as
\begin{eqnarray}
F'^{\mu \nu}(x') = \frac{\6 x'^{\mu}}{\6 x^{\lambda}}\frac{\6 x'^{\nu}}{\6 x^{\rho}} F^{\lambda \rho}(x) 
\label{def1}
\end{eqnarray}
Boost transformations are written as 
\begin{equation}
    x'^0 = \gamma x^0 - \frac{\gamma v_i}{c} x^i
    \label{t6}
\end{equation}
\begin{equation}
    x'^i= x^i - \frac{\gamma v^i}{c}x^0 + (\gamma -1 )\frac{v^iv_j}{v^2}x^j
    \label{t7}
\end{equation}
From \ref{def1} using \ref{t6} and \ref{t7} we get following relations
\begin{eqnarray}
F'^{0i} = E'^i = \gamma E^i + \frac{\gamma-1}{v^2} v^i v_j E^j - \frac{\gamma v_j}{c} F^{ji} \label{e}
\end{eqnarray}
\begin{eqnarray}
F'^{ij} 
=  -\frac{\gamma v^i}{c} E^j+ \frac{\gamma v^j}{c} E^i + F^{ij} + \frac{\gamma - 1 }{v^2}v_m v^j F^{im} + \frac{\gamma -1 }{v^2} v_l v^i F^{lj}\label{b}
\end{eqnarray}
\noindent \underline{\bf Electric limit}\\
\noindent From eqn \ref{e} using the electric limit scaling given in table \ref{T1} and keeping in mind that in this limit $\gamma \to 1$ as $c \to \infty$, we get
\begin{eqnarray}
 e'^i = e^i \label{114}
\end{eqnarray}
\noindent Similarly, eqn \ref{b} yields
\begin{eqnarray}
 f'^{ij} =- v^i e^j + v^j e^i + f^{ij} 
\implies b'^k =  b^k -\Big(\vec v \times \vec e \Big)^k \label{115}
\end{eqnarray}
The transformations \ref{114} \& \ref{115} manifest the same asymmetry that was observed in the Maxwell's equations \ref{max1}, \ref{max2}. A change in the electric field induces a change in the magnetic field but the converse is not true. For the magnetic limit, discussed right below, it is the other way round.\\
\indent There is a simple group theoretical argument for the absence of any $b$-term in \ref{114}. For argument's sake if we retain \ref{115} but include a term in \ref{114} like 
\begin{equation}
   \vec e' = \vec e + \Big( \vec v \times \vec b \Big)  \label{p}
\end{equation}
then the group composition law fails since, 
\begin{equation}
    \vec e'' = \vec e' + \Big( u \times \vec b' \Big) = \vec e + \Big( (\vec v + \vec u) \times \vec b \Big) - \vec u \times (\vec v \times \vec e)
\end{equation}
and the last term spoils the transformation \ref{p}.\\
\noindent \underline{\bf Magnetic limit}\\
From eqn \ref{e} using the magnetic limit scaling from table \ref{T1} we get
\begin{eqnarray}
 e'^i = e^i - v_j f^{ji} 
\implies e'^i = e^i + \Big(\vec v \times \vec b \Big)^i \label{econd}
\end{eqnarray}
Similarly from eqn \ref{b} we get
\begin{eqnarray}
 f'^{ij} = f^{ij} 
\implies \vec b' = \vec b \label{bcond}
\end{eqnarray}
Adopting the same method the transformations in the covariant sector are obtained. All these results are summarised in table \ref{T04}. 
\begin{table}
\caption{Transformation of fields under Galilean boost}\label{T04}
\begin{center}
\begin{tabular}{|c|c|c|} \hline 
${\rm Limits}$  & $ {\rm Contravariant ~case}$ & $ {\rm Covariant ~case}$  \\ \hline
${\rm Electric ~limit}$ & $e'^i = e^i, ~~b'^k =  b^k -\Big(\vec v \times \vec e \Big)^k$ & $e'_i = e_i + \Big(\vec v \times \vec b  \Big)_i, ~~\vec b' = \vec b$\\ \hline
${\rm Magnetic ~limit }$ &  $e'^i = e^i + \Big(\vec v \times \vec b \Big)^i, ~~\vec b' = \vec b $  & $e'_i = e_i, ~~b'_k = b_k - \Big(\vec v \times \vec e \Big)_k$ \\
\hline
\end{tabular}
\label{T04}
\end{center}
\end{table}
We can clearly see from this table that under duality transformation ($e^i \to b^i, ~b^i \to -e^i$ and $e_i \to b_i, ~b_i \to -e_i$) electric limit reproduces magnetic limit and vice-versa for both covariant and contravariant cases.
\section{Gauge symmetry} \label{sec5}
We know in the relativistic case the Maxwell lagrangian,
\begin{equation}
  \mathcal{L} = -\frac{1}{4} F_{\mu \nu} F^{\mu \nu}  
\end{equation}
is invariant under the following gauge transformation,
\begin{equation}
    \delta A_{\mu} = \6_{\mu} \alpha, \,\,\,  \delta A^{\mu} = \6^{\mu} \alpha \label{cpot}
\end{equation}


\noindent We consider the Galilean version of this gauge invarianvce. \\
 \subsection{Galilean version} \label{gv}
Here we can consider a relatively more general gauge condition,
\begin{equation}
    \delta A_{\mu} = \6_{\mu} \alpha, \,\,\,  \delta A^{\mu} = \6^{\mu} \beta \label{cpot1}
\end{equation}

\noindent In the relativistic theory the covaraint and contravariant vectors are related by a metric implying $\alpha = \beta$. This is mot true in galilean limit. Hence we take $\alpha \neq \beta$ when deriving the galilean version of the gauge transformations. First we consider the electric limit. \\
\noindent \underline {\bf Electric limit} \\
From eqn \ref{cpot1} and using the mapping relations given in table \ref{T1}  we deduce the following relations,
\begin{eqnarray}
\delta A_0 = \6_0 \alpha \implies \frac{1}{c} \delta a_0 = \frac{1}{c} \6_t \alpha 
\implies \delta a_0 = \6_t \alpha \label{eg1}
\end{eqnarray}
\begin{eqnarray}
\delta A_i = \6_i \alpha \implies \delta a_i = \6_i \alpha \label{eg2}
\end{eqnarray}
\begin{eqnarray}
\delta A^0 = \6^0 \beta \implies c \delta a^0 = -\frac{1}{c} \6_t \beta   
\xrightarrow[\text{$c \to \infty$}]{\text{}} \delta a^0 = 0 \label{eg3}
\end{eqnarray}
\begin{eqnarray}
\delta A^i = \6^i \beta \implies \delta a^i = \6^i \beta \label{eg4}
\end{eqnarray}

\noindent Taking the variation of \ref{el1} in the electric limit,
\begin{eqnarray}
\delta \mathcal{L}_e = \frac{1}{2} \6^i \delta a^0 \Big(\6_t a_i - \6_i a_0  \Big) + \frac{1}{2} \6^i a^0 \Big(\6_t \delta a_i - \6_i \delta a_0  \Big)  = 0 \label{clag}
\end{eqnarray}
on exploiting eqns \ref{eg1}, \ref{eg2}, \ref{eg3}, \ref{eg4}. This shows the invariance of $\mathcal{L}_e$.\\
\noindent \underline {\bf Magnetic limit} \\
From eqn \ref{cpot1} and using the mapping relations given in table \ref{T1} and repeating the steps done for the electric limit, we obtain similar results here also. These are given in table \ref{T3}. 

Taking the variation of the lagrangian \ref{l2} and using the results in table \ref{T3} we get
\begin{eqnarray}
 \delta \mathcal{L}_m = \frac{1}{2} \6_i \delta a_0 \Big(\6_t a^i - \6^i a^0 \Big) + \frac{1}{2} \6_i a_0 \Big(\6_t \delta a^i - \6^i \delta a^0 \Big) 
 = 0 
\end{eqnarray}
 This shows the invariance of $\mathcal{L}_m$.
\begin{table}
\caption{Variations of the Galilean potentials}\label{T3}
\begin{center}
\begin{tabular}{|c|c|c|} \hline 
${\rm Variable}$  & $ {\rm Electric ~~limit}$ & $ {\rm Magnetic ~~limit}$  \\ \hline
$a^0$ & $\delta a^0 = 0$ & $\delta a^0 = \6_t \beta$\\ \hline
$a^i$ & $\delta a^i = \6^i \beta$  & $\delta a^i = \6^i \beta$ \\ \hline
$a_0$ & $\delta a_0 = \6_t \alpha$ & $\delta a_0 = 0 $\\ \hline
$a_i$ & $\delta a_i = \6_i \alpha$  & $\delta a_i = \6_i \alpha$\\ 
\hline
\end{tabular}
\label{T3}
\end{center}
\end{table}
\subsection{Noether current conservation}
\noindent We know in relativistic classical field theory the Noether current is defined as
\begin{eqnarray}
 J^{\mu}= \frac{\6 \mathcal{L}}{\6 (\6_\mu A_\nu)} \delta A_\nu \label{current1}
\end{eqnarray}
which is conserved on-shell i.e $\6_\mu J^\mu =  0$. Specifically for the 
Maxwell Lagrangian,
\begin{equation*}
  \mathcal{L} = -\frac{1}{4} F_{\mu \nu} F^{\mu \nu}  
\end{equation*}
the current in eqn \ref{current1} has the form, 
\begin{equation}
    J^\mu = -F^{\mu \nu} ~\6_\nu \alpha 
\end{equation}
and is on-shell conserved i.e
\begin{eqnarray}
   \6_\mu J^\mu = -\Big(\6_\mu F^{\mu \nu} \Big)\6_\nu \alpha - F^{\mu \nu} \6_\mu \6_\nu \alpha 
   =0
\end{eqnarray}
The first term is zero because $\6_\mu F^{\mu \nu}= 0 $ and second term is zero because of anti-symmetry of $F^{\mu \nu}$.\\
\indent We now consider here a suitable Galilean version of this conservation. For this we will directly start from the relativistic definition and substitute the Galilean results in proper limit (electric or magnetic). \footnote{The details of the Noether current calculations are provided in appendix \ref{ap1}.} Analogous conservation laws, either in electric or magnetic limit are obtained.

\section{Shift symmetry} \label{sec6}
We know that Goldstone's theorem is a crucial input of the study of low-energy effective lagrangians implying that whenever a global symmetry is spontaneously broken, a gapless mode will appear. In relativistic theories this leads to a massless Goldstone particle described by a shift symmetry of
the field
\begin{equation}
    \phi(x) \to \phi(x) + c \label{fi}
\end{equation}
where $c$ is constant and is characterised by the scalar field action
\begin{equation}
    S = \frac{1}{2} \int d^d x \6_\mu \phi \6^\mu \phi
\end{equation}
The above action is invariant under \ref{fi}. Since \ref{fi} is a global transformation, the conserved currents can be found by exploiting Noether's first theorem 
\begin{eqnarray}
J^\mu = \frac{\6 \mathcal{L}}{\6(\6_\mu \phi)} \delta \phi = c ~\6^\mu \phi
\end{eqnarray}
And corresponding conservations are demonstrated as 
\begin{equation}
    \6_\mu J^\mu = c \6_\mu \6^\mu \phi = 0
\end{equation}
Consider a constant shift in the four potential, 
\begin{equation}
    A'_{\mu} = A_{\mu} + C_{\mu}, \, \, \, \, A'^{\mu} = A^{\mu} + D^{\mu}
\end{equation}
that leaves Maxwell lagrangian invariant. We take $C$ and $D$ to be different for reasons stated in section \ref{gv}.  \\
\noindent \underline {\bf Electric limit} \\
We can define following things
\begin{eqnarray}
 \delta A_0 = C_0 \implies \frac{1}{c} \delta a_0 = C_0 \implies \delta a_0 = c C_0 \\
\delta A_i = C_i \implies \delta a_i = C_i 
\end{eqnarray}
Similarly, from expressions for $\delta A^0, ~\delta A^i$ we find,
\begin{eqnarray}
  \delta a^0 =  0, \,\,\,~~~ \delta a^i = D^i 
\end{eqnarray}
From \ref{el1} Noether currents are found to be
\begin{eqnarray}
j^t = \frac{1}{2} (\6^i a^0) C_i, \,\,\, ~~~~~j^i =  -\frac{1}{2} \6^i a^0 C_0 - \frac{1}{2} f^{ij} C_j
\end{eqnarray}
And current conservation can be explicitly demonstrated as 
\begin{eqnarray}
\6_\mu J^\mu  \xrightarrow[\text{$c \to \infty$}]{\text{Electric limit}}\6_t j^t + \6_i j^i = \frac{1}{2} (\6_t \6^i a^0) C_i - \frac{1}{2} (\6_i \6^i a^0) C_0 - \frac{1}{2} (\6_i f^{ij}) C_j 
= 0 
\end{eqnarray}
The covariant components of the currents are
\begin{eqnarray}
j_t = 0, \,\,\, ~~~~~j_i = - \frac{1}{2} f_{ij} D^j 
\end{eqnarray}
The current conservation gives us 
\begin{eqnarray}
\6^\mu J_\mu  \xrightarrow[\text{$c \to \infty$}]{\text{Electric limit}}\6^i j_i = -\frac{1}{2} (\6^i f_{ij}) D^j = 0 
\end{eqnarray}
\noindent \underline {\bf Magnetic limit} \\
We can define following things
\begin{eqnarray}
\delta A_0 = C_0 \implies -c \delta a_0 = C_0 \implies \delta a_0 = 0 \\
\delta A_i = C_i \implies \delta a_i = C_i 
\end{eqnarray}
Similarly, from expressions for $\delta A^0, ~\delta A^i$ we find,
\begin{eqnarray}
 \delta a^0 =  -c D^0, \,\,\, ~~~~~~ \delta a^i = D^i 
\end{eqnarray}
From \ref{l2} the Noether currents are found to be 
\begin{eqnarray}
j^t =0, \,\,\, ~~~~ j^i =-\frac{1}{2} f^{ij} C_j
\end{eqnarray}
So the current conseravtaions are demonstrated as
\begin{eqnarray}
\6_\mu J^\mu  \xrightarrow[\text{$c \to \infty$}]{\text{Magnetic limit}} \6_i j^i = -\frac{1}{2} \6_i f^{ij} C_j = 0
\end{eqnarray}
Similarly the covariant current components are 
\begin{eqnarray}
j_t = \frac{1}{2} \6_i a_0 D^i, \,\,\, ~~~~~ j_i = -\frac{1}{2} \6_i a_0 D^0 - \frac{1}{2} f_{ij} D^j
\end{eqnarray}
The current conservations give
\begin{eqnarray}
\6^\mu J_\mu  \xrightarrow[\text{$c \to \infty$}]{\text{Magnetic limit}}\6_t j_t + \6^i j_i 
= \frac{1}{2} (\6_t \6_i a_0) D^i - \frac{1}{2} (\6^i \6_i a_0) D^0 - \frac{1}{2}\6^i f_{ij} D^j = 0 
\end{eqnarray}

\section{Inclusion of sources} \label{sec7}
The relativistic Maxwell Lagrangian with source is as follows
\begin{equation}
    \mathcal{L} = -\frac{1}{4} F_{\alpha \beta} F^{\alpha \beta} - A_{\alpha} J^{\alpha}
\end{equation}
We can write the Lagrangian in the following form for convenience
\begin{equation}
    \mathcal{L} = -\frac{1}{4} \Big(2 F_{0i}F^{0i} + F_{ij} F^{ij} \Big) - \frac{1}{2} A_{\a} J^{\a} - \frac{1}{2} A^{\a} J_{\a}
\end{equation}
We know that the Maxwell theory respects the following gauge transformations
\begin{equation}
    A_\mu \to A_\mu + \6_\mu \Lambda
\end{equation}
The gauge invariance of the action demands the following condition
\begin{equation}
  A_\mu J^\mu \to  \Big(A_\mu + \6_\mu \Lambda \Big) J^\mu = A_\mu J^\mu - \Lambda \6_\mu J^\mu \implies \6_\mu J^\mu = 0 
\end{equation}
\noindent \underline {\bf Electric limit:}\\
\noindent In the electric limit the scaling of the components of the source $J^\mu$ will be as follows
\begin{eqnarray}
 J^0 \to c j^0, \,\, J^i \to j^i, \,\, J_0 \to \frac{j_0}{c}, \,\, J_i \to j_i
 \label{jmapping}
\end{eqnarray}
In this limit the Lagrangian looks like 
\begin{equation}
    \mathcal{L}_e = \frac{1}{2} \6^i a^0 \Big(\6_t a_i - \6_i a_0  \Big) -\frac{1}{4} f_{ij} f^{ij} - \frac{1}{2} a_0 j^0 -\frac{1}{2} a_i j^i - \frac{1}{2} a^0 j_0 -\frac{1}{2} a^i j_i
\label{ls1}    
\end{equation}
Varying the lagrangian with respect to $a_0, ~a_j, ~a^0, ~a^j$ will give following set of equations
\begin{eqnarray}
 \6^i \6_i a^0 = j^0
 \label{es1}
\end{eqnarray}
\begin{eqnarray}
 \6_t \6^j a^0 + \6_i \6^j a^i - \6_i \6^i a^j = - j^j \label{es2}
\end{eqnarray}
\begin{eqnarray}
\6^i \6_t a_i - \6^i \6_i a_0 = -j_0 \label{es3}
\end{eqnarray}
\begin{eqnarray}
\6^i \6_i a_j - \6^i \6_j a_i = j_j \label{es4}
\end{eqnarray}
We now derive these equations directly from the equations of motion. The relativistic equations are given by,
\begin{eqnarray}
\6_i F^{i0} =J^0, \,\,\,\,\,~~~~~\6_0 F^{0j} + \6_i F^{ij} = J^j \label{ems2}
\end{eqnarray}
Using maps given in table \ref{T1} and \ref{jmapping} it is simple to verify that they reproduce eqns \ref{es1} and \ref{es2}. Using the covariant counterpart of eqns \ref{ems2} we can get eqns \ref{es3} and \ref{es4}. 
This shows the consistency of the eqn of motion in Galilean electrodynamics with source.\\
 Taking the variation of the source part of the lagrangian we get 
 \begin{eqnarray}
 \delta \mathcal{L}_e = -\frac{1}{2} \6_t \alpha j^0 - \frac{1}{2} \6_i \alpha j^i - \frac{1}{2} \6^i \beta j_i = \frac{1}{2}\alpha \Big(\6_t j^0 + \6_i j^i \Big) + \frac{1}{2} \beta \6^i j_i = 0
 \end{eqnarray}
Since $\alpha, \beta \neq 0$ we have two conditions
\begin{eqnarray}
\6_t j^0 + \6_i j^i = 0, \, \, \, \, \, \6^i j_i = 0
\end{eqnarray}
The sources as given in eqns (\ref{es1})-(\ref{es4}) satisfy the above conditions. We observe that sources are conserved off-shell.\\
\noindent \underline {\bf Magnetic limit:}\\
Here the scaling relations are as follows
\begin{eqnarray}
 J^0 \to -\frac{j^0}{c}, \,\, J^i \to j^i, \,\, J_0 \to -c j_0, \,\, J_i \to j_i 
\end{eqnarray}
The Lagrangian in this limit is as follows 
\begin{equation}
    \mathcal{L}_m = \frac{1}{2} \6_i a_0 \Big(\6_t a^i - \6^i a^0 \Big) - \frac{1}{4} f_{ij}f^{ij} - \frac{1}{2} a_0 j^0 -\frac{1}{2} a_i j^i - \frac{1}{2} a^0 j_0 -\frac{1}{2} a^i j_i
\end{equation}
Varying the lagrangian wrt $a_0, ~a_j, ~a^0, ~a^j$ we get the following set of equations
\begin{eqnarray}
 \6_i \6_t a^i - \6_i \6^i a^0 = -j^0 \label{ms1}
\end{eqnarray}
\begin{eqnarray}
 \6_i \6^i a^j - \6_i \6^j a^i = j^j \label{ms2}
\end{eqnarray}
\begin{eqnarray}
\6^i \6_i a_0 = j_0 \label{ms3}
\end{eqnarray}
\begin{eqnarray}
\6_t \6_j a_0 - \6^i \6_i a_j + \6^i \6_j a_i = - j_j \label{ms4}
\end{eqnarray}
These equations may also be derived directly from the equations of motion following the same method adopted for the electric limit. The field equations for both electric and magnetic limit have been shown in table \ref{T7}.
\begin{table}
\caption{Field equations}\label{T7}
\begin{center}
\begin{tabular}{|c|c|c|} \hline 
${\rm Variables}$ & ${\rm Electric ~~limit}$ & ${\rm Magnetic ~~limit}$ \\ \hline
$a^0$  & $\6^i \6_t a_i - \6^i \6_i a_0 = -j^0$ & $\6^i \6_i a_0 = j^0$ \\ \hline
$a^j$ & $\6^i \6_i a_j - \6^i \6_j a_i = j_j $ & $\6_t \6_j a_0 + \6^i \6_j a_i - \6^i \6_i a_j = -j_j$ \\ \hline
$a_0$ & $ \6^i \6_i a^0 = j^0 $ & $\6_i \6_t a^i - \6^i ~\6_i ~a^0 = -j^0 $ \\ \hline
$a_j$ & $\6_t \6^j a^0 + \6_i \6^j a ^i - \6_i \6^i a^j = -j^j$ & 
$\6_i\6^i a^j - \6_i \6^j a^i = j^j $ \\ \hline
\end{tabular}
\label{T7}
\end{center}
\end{table}
\noindent Taking the variation of the source part of the lagrangian we get
\begin{equation}
    \delta \mathcal{L}_m = -\frac{1}{2} \6_i \alpha j^i - \frac{1}{2} \6_t \beta j_0 - \frac{1}{2} \6^i \beta j_i = \frac{1}{2}\alpha \Big(\6_i j^i \Big) + \frac{1}{2} \beta \Big(\6_t j_0 + 6^i j_i \Big) = 0
\end{equation}
Since $\alpha, \beta \neq 0$ we have two conditions
\begin{eqnarray}
\6_i j^i = 0, \, \, \, \, \6_t j_0 + \6^i j_i = 0 
\end{eqnarray}
Here also sources given by eqns (\ref{ms1}) to (\ref{ms4}) satisfy the above off-shell conservation equations. 
\section{Conclusions} \label{sec8}
\noindent Let us summarise, point by point, the new significant findings of the paper, comparing with existing results found in the literature.
\begin{itemize}
    \item An unambiguous construction of the non-relativistic (NR) lagrangian, for both electric and magnetic limits, was given. We have shown that it correctly reproduces the equations of motion either in the potential or field (electric/magnetic) formulation. This lagrangian was deduced from the standard relativistic lagrangian adopting the dictionary given here. It is expressed either in terms of potentials or fields. In the later case both electric and magnetic limit lagrangians become identical having the same functional form as the usual Maxwell lagrangian.\\ In ref \cite{Mehra3}, a NR lagrangian has been given, also derived from the relativistic Maxwell expression, which has, however, several shortcomings. \footnote{Details are provided in the Appendix \ref{ap2}} 
    \item It is observed from table \ref{T2} that if we replace the covariant components by contravariant ones in the electric limit case we will end up with the magnetic limit case and vice-versa. This fact manifests itself only if we consider the covariant and contravariant sectors separately as we have done here.The interplay between the covariant and the contravaraint indices that leads to an interchange of the electric and the magnetic limits of the theory is a new feature observed here. The reason that it was not noticed earlier stems from the fact that various applications \cite{Leblond, Khanna, Duval, Mehra1,  Bleeken, Bergshoeff, Festuccia, Mehra3, Chapman, Sharma} only considered the contravariant components. There is a paper \cite{St} that only gives the Galilean transformation for both covariant and contravariant components and that too confined to the coordinates and derivatives, and not for an arbitrary field. Our analysis is much more general where we provide maps, for both covariant and contravariant sectors, relating arbitrary four vectors in the Lorentz relativistic and Galilean relativistic formulations. These maps are the genesis of our analysis where we use them to obtain galilean relativistic expressions from their corresponding Lorentz relativistic counterparts.  These issues are not even remotely mentioned, much less discussed, in \cite{St}.
    \item A central point is the formulation of a dictionary that translates four vectors in the relativistic theory to their corresponding vectors in the non-relativistic theory. Thus the formalism developed in terms of potentials  was extended to field (electric \& magnetic) formulation. In this set up the duality symmetry was discussed. One can clearly see that in the non-relativistic limit the duality relations are quite non-trivial. In this limit we show that apart from the usual duality relations a twisted duality relation also exists. The feature of twisted duality manifests precisely because the covariant and contravariant vectors are treated separately. This also shows that, on the lagrangian level, duality plays quite a subtle role.\\ Duality symmetries have useful physical  applications. For standard Maxwell's theory, using duality symmetry we can find new solutions from given original solutions. Here duality symmetry switches from the electric limit to the magnetic limit. Thus the solutions of the Rowland-Vasilescu Karpen's effect which is an example of the Galilean electric limit, can be exploited, using the duality relations, to find solutions of Wilson's effect which corresponds to the Galilean magnetic limit.\footnote{These effects and their implications have been discussed in \cite{Germain}, but there is no mention of duality symmetry.}. 
    \item Gauge symmetries play a pivotal role in the understanding of gauge theories. Since covariant ($a_\mu$) and the contravarinat ($a^\mu$) vectors are not connected by any non-degenerate metric, they have separate gauge transformations. While this was noticed earlier \cite{Mehra3}, its full implications were not analysed, and not just because of their problematic lagrangian \ref{apl1}. We show how gauge symmetries in the relativistic case naturally yield their non-relativistic counterpart, but with distinct gauge parameters. Both electric and magnetic limits were analysed. The conservation laws were derived using Noether's prescription. 
    \item Shift symmetries, which have an important role to describe Goldstone particles in relativistic theories, were introduced in the non-relativistic context. Conservation laws, associated with such symmetries, were derived in both electric and magnetic limits. 
    \item We have provided a completely holistic approach in terms of both potentials and fields, clearly showing the connection among them, starting from the rudimentary structures of usual Maxwell's theory. Such an analysis is lacking in the literature. 
\end{itemize}
\noindent \underline{\bf Future prospects:}\\
\noindent This is quite a new research area and  has gained
attention of late as a part of the resurgence of non-Lorentzian structures in quantum field theories, holography and string theory and hence many aspects and directions are yet to be looked at. There is no consistent Hamiltonian formalism for galilean electrodynamics for example.  Also it will be interesting to study the non-relativistic limits of other gauge theories for example Proca theory which describes a massive spin 1 field or Maxwell Chern-Simons theory which is a $2+1$ dimensional gauge theory, in the same way described here. The analysis described here for vector field could be extended to include tensor fields like the Kalb-Ramond fields. Since the connection of these fields with  non-relativistic fluid dynamics is known \cite{Nambu, Sugamoto}, though relatively less studied, the present formulation could find application to illuminate this connection. All these possibilities should be tractable since we have provided independent maps for both covariant and contravariant sectors. Finally, we hope to elucidate the nature of Carrollian electrodynamics \cite{Duval} using the methods developed here. We expect we can address these issues in the near future. 
\section{Acknowledgements}
The authors (RB and SB) acknowledge the support from a DAE Raja Ramanna Fellowship (grant no: \\$1003/(6)/2021/RRF/R\&D-II/4031$, dated: $20/03/2021$). They also acknowledge Sumit Dey for useful discussions. 
\begin{appendices}
\section{Noether current calculation} \label{ap1}
\noindent \underline {\bf Electric limit} \\
The contravariant components of the current for the relativistic case are
\begin{eqnarray}
J^0 = \frac{\6 \mathcal{L}}{\6 (\6_0 A_\nu)}\delta A_\nu \label{a1}
\end{eqnarray}
\begin{eqnarray}
J^i = \frac{\6 \mathcal{L}}{\6 (\6_i A_0)}\delta A_0 + \frac{\6 \mathcal{L}}{\6 (\6_i A_j)}\delta A_j \label{a2}
\end{eqnarray}
Now using the maps for electric limit given in table \ref{T1} we get
\begin{eqnarray}
 c j^0 = c \frac{\6 \mathcal{L}_e}{\6 (\6_t a_0)}\delta a_0  + c \frac{\6 \mathcal{L}_e}{\6 (\6_t a_i)}\delta a_i 
\implies j^0 = \frac{1}{2} \6^i a^0 \6_i \alpha 
\end{eqnarray}
Similarly, 
\begin{eqnarray}
 j^i = \frac{\6 \mathcal{L}_e}{\6 (\6_i a_0)}\delta a_0 + \frac{\6 \mathcal{L}_e}{\6 (\6_i a_j)}\delta a_j 
= -\frac{1}{2} \6^i a^0 \6_t \alpha - \frac{1}{2} \Big( \6^i a^j - \6^j a^i\Big) \6_j \alpha
\end{eqnarray}
So we can show the conservation of the galilean currents as follows 
\begin{eqnarray}
\6_\mu J^\mu  \xrightarrow[\text{$c \to \infty$}]{\text{Electric limit}}  \6_t j^0 + \6_i j^i 
= \frac{1}{2} \6_t \6^i a^0 \6_i \alpha +  \6_i [-\frac{1}{2} \6^i a^0 \6_t \alpha - \frac{1}{2} \Big( \6^i a^j - \6^j a^i\Big) \6_j \alpha] \nonumber \\
= -\frac{1}{2} \Big(\6_i \6^i a^0 \Big) \6_t \alpha + \frac{1}{2}\Big(\6_t \6^j a^0 - \6_i f^{ij} \Big) \6_j \alpha + \frac{1}{2} \6^i a^0 \6_t \6_i \alpha - \frac{1}{2} \6^i a^0 \6_i \6_t \alpha - \frac{1}{2} f^{ij} \6_i \6_j \alpha = 0 \label{conservn1}
\end{eqnarray}
In the second line of eqn \ref{conservn1}, the first and second term is zero from equations of motion eqn \ref{ee1} and \ref{ee2} respectively, third and fourth terms get cancelled and fifth term vanishes because of antisymmetry. \\
Following identical arguments current conservation for covariant components can be shown as
\begin{eqnarray}
\6^\mu J_\mu  \xrightarrow[\text{$c \to \infty$}]{\text{Electric limit}}  \6^i j_i 
=  0, \label{conservn2}
\end{eqnarray}
\noindent \underline {\bf Magnetic limit} \\
Using the maps for magnetic limit given in table \ref{T1} and exploiting eqns \ref{a1} and \ref{a2} we get
\begin{eqnarray}
 j^0 
= -c^2 \frac{\6 \mathcal{L}_m}{(-c)\6 (\6_t a_0)}\delta a_0 -\frac{\6 \mathcal{L}_m}{\6 (\6_t a_i)}\delta a_i 
= 0 
\end{eqnarray}
\begin{eqnarray}
 j^i = 
= \frac{\6 \mathcal{L}_m}{(-c)\6 (\6_i a_0)}(-c)\delta a_0 + \frac{\6 \mathcal{L}_m}{\6 (\6_i a_j)}\delta a_j 
= -\frac{1}{2} f^{ij} \6_j \alpha 
\end{eqnarray}
Using these expressions, 
\begin{eqnarray}
\6_\mu J^\mu  \xrightarrow[\text{$c \to \infty$}]{\text{Magnetic limit}} \6_i j^i 
= -\frac{1}{2} \Big(\6_i \6^i a^j - \6_i \6^j a^i \Big) \6_j \alpha = 0 \label{conservn3}
\end{eqnarray}
where the second equality vanishes from eqn \ref{em2}.
Likewise for the covariant case, following identical arguments current conservation can be shown as
\begin{eqnarray}
\6^\mu J_\mu  \xrightarrow[\text{$c \to \infty$}]{\text{Magnetic limit}}\6_t j_0 + \6^i j_i
=0 \label{conservn4}
\end{eqnarray}
\section{Problems and inconsistencies of the lagrangian formulation given in \cite{Mehra3}} \label{ap2}
\noindent Any consistent lagrangian formulation of galilean electrodynamics must yield all the equations of motion, for either contravariant or covariant vectors in both electric and magnetic limits. Simultaneously, these equations must reduce to those given in \cite{Leblond} using the field (electric/magnetic) formulation. This is not merely desirable, but essential, since those equations were obtained directly \cite{Leblond} using galilean relativistic arguments, bypassing the use of limiting prescriptions. Since the basic variables in the lagrangian are the potentials, equations of motion are obtained in the potential formulation. One has to now express the electric and magnetic fields in terms of potentials and recast the equations of motion in terms of these fields. Only then a comparison with \cite{Leblond} is feasible. As we explicitly show, the lagrangian given in \cite{Mehra3} fails on all counts. \\ 
\indent The covariant Galilean relativistic largrangian given in \cite{Mehra3} is 
\begin{equation}
        \mathcal{L} = -\frac{1}{4} \Big( \6_\mu a_\nu - \6_\nu a_\mu \Big) \Big( \6^\mu a^\nu - \6^\nu a^\mu \Big) \label{apl1}
    \end{equation}
    which gives rise to following set of equations
    \begin{equation}
        \6_\mu \Big( \6^\mu a^\nu - \6^\nu a^\mu \Big) =0, \label{apee1} \,\,\,\,\,\,\,~~~~~ ({\rm Electric ~~limit})
    \end{equation}
    \begin{equation}
      \6^\mu \Big(\6_\mu a_\nu - \6_\nu a_\mu \Big) = 0, \label{apme1} \,\,\,\,\,\,\,~~~~~ ({\rm Magnetic ~~limit})
    \end{equation}
   \noindent Let us first consider eqn \ref{apee1}. If we re-write this equation component wise it gives the following set of equations \footnote{According to the convention used in \cite{Mehra3}, $\6^\mu = (0,~\6^i)$ and $\6_\mu = (\6_t, ~\6_i)$}.
   \begin{eqnarray}
       \6_i \6^i a^0 = 0, \,\,\,\,\,~~~~~~~ \6_t \6^i a^0 + \6_j \6^i a^j - \6_j \6^j a^i =0 
   \end{eqnarray}
   which are nothing but equation \ref{ee1} and \ref{ee2} respectively. But we cannot get eqns \ref{ee3} and \ref{ee4} from the lagrangian \ref{apl1}. In fact we cannot get any equation involving $a_0$ and / or $a_i$ simply because there are no covariant components.\\
   \indent Similarly we can open the magnetic limit equation \ref{apme1} component wise as follows
   \begin{eqnarray}
       \6^i \6_t a_i - \6^i \6_i a_0 = 0, \,\,\,\,\, ~~~~~\6^j \6_i a_j - \6^j \6_j a_i = 0 \label{apme2}
   \end{eqnarray}
   These two equations do not correspond to any of our equations. On top of that there are no equations for $a^0$ and / or $a^i$ simply because contravariant indices do not arise.\\ 
   \indent Thus the lagrangian \ref{apl1} fails to yield, in the electric limit, any equation involving covariant indices for potentials. Likewise, in the magnetic limit, there are no equations in the contravariant sector. It is also not possible to express equations (\ref{apee1}, \ref{apme1}) in the field formulation since no map ralating potentials with fields has been given. Hence the mandatory comparison with \cite{Leblond} cannot be done. All these issues have been discussed successfully in our approach.\\
   \indent If we push the analysis of \cite{Mehra3} further, serious inconsistencies arise. The master equations \ref{apme2}, from which the lagrangian \ref{apl1} was written, was claimed to be derived by opening the relativistic Maxwell equation \ref{e2}  in space-time components and exploiting the map given in \cite{Mehra3},
   \begin{equation}
       A^0 = -\frac{1}{c} a_0,  \,\,\,\,\,\,\,\, A^i = a_i
   \end{equation}
   Doing this, however, instead of \ref{apme2} we find
   \begin{equation}
       \6_i \6^i a_0 = 0, \,\,\,\,\,\,\,\, \6_j \6^i a_j - \6_j \6^j a_i = 0 
   \end{equation}
   Surprisingly, there is a mismatch with the first equality in \ref{apme2}. Thus the very construction of the lagrangian and the associated equations of motion in \cite{Mehra3} are all riddled with inconsistencies.
\end{appendices}

\end{document}